\documentclass[a4paper,11pt]{article}
\usepackage{pos}
\usepackage{lmodern}
\usepackage{todonotes}
\usepackage{subcaption}

\title{
  Applying template synthesis to the radio emission
  from air showers with generic geometries
}
\ShortTitle{Template synthesis for general geometries}

\author*[a]{Mitja Desmet}
\author[a]{Stijn Buitink}
\author[a,b]{Tim Huege}

\affiliation[a]{
  Inter-University Institute For High Energies (IIHE), Vrije Universiteit Brussel (VUB),\\
  Pleinlaan 2, 1050 Brussels. Belgium
}

\affiliation[b]{
  Institute for Astroparticle Physics (IAP), Karlsruhe Institute of Technology (KIT),\\
  PO Box 3640, 76021 Karlsruhe, Germany
}

\emailAdd{mitja.desmet@vub.be}

\abstract{
  Studying high-energy cosmic-ray air showers through the radio emission produced by their secondary particles is a well-established technique. 
  However, due to the increasing size and density of the radio arrays, analyses are running into computational limits, as these rely on Monte Carlo simulations to model the emission.
  To address this, we have been developing template synthesis, a new approach to simulate the radio emission from air showers. 
  With template synthesis, we use semi-analytical expressions to describe how the radio emission from an air shower depends on the shower age and the position of the antenna with respect to the shower. 
  These expressions are extracted from a set of microscopic simulations, thus benefiting from their accuracy. 
  Once obtained, we can use these relations to synthesise the emission from an air shower with any longitudinal profile, by using a single Monte Carlo simulation as an input.
  Previously we have demonstrated that this hybrid approach can synthesise the radio emission from air showers and agrees with results from microscopic simulations within 10\%. 
  The method was however limited to a specific geometry.
  Here we present our first step towards generalising template synthesis across geometries. 
  We found a set of scaling relations which correct for the shower geometry as well as the viewing angle under which the radiation is observed. 
  This allows us to reformulate the semi-analytical relations in a way that does not longer depend on the geometry, significantly reducing the number of parameters that need to be fitted.
  We apply these scaling relations to a simulation library of CORSIKA showers with a zenith angle of 50 degrees, with primary energies between $10^{17} \; \text{eV}$ and $10^{19} \; \text{eV}$. 
  We then extract the semi-analytical expressions required for template synthesis, and use them to synthesise the emission from air showers with lower zenith angles. 
  We investigate the accuracy by comparing both to microscopic simulations as well as the single geometry version of template synthesis.
}

\FullConference{
  10th International Workshop on Acoustic and Radio EeV Neutrino Detection Activities (ARENA2024)\\
  11-14 June 2024\\
  The Kavli Institute for Cosmological Physics, Chicago, IL, USA\\
}

\renewcommand{\vec}[1]{\boldsymbol{#1}}

\newcommand{\Xmax}{X_{\text{max}}}
\newcommand{\textXmax}{$X_{\text{max}}$}

\newcommand{\textgcm}{$\text{g} \, \text{cm}^{-2}$}
\newcommand{\slice}[1]{{#1}_{\text{slice}}}
\newcommand{\DeltaXmax}{\Delta \Xmax}
\newcommand{\textDeltaXmax}{$\DeltaXmax$}

\begin{document}
\maketitle

\section{Introduction}
\label{sec:intro}

Detecting the radio emission from the extensive air showers (EAS) produced by cosmic rays entering our atmosphere, is an attractive way to instrument the large areas required to measure their flux at the highest energies. 
The emission from this cascade of secondary particles is predominantly produced by the electrons and positrons. 
In air, we can distinguish two main macroscopic emission components: the geomagnetic and charge-excess (also referred to as Askaryan) mechanisms \cite{Huege2016}.
The first one is dominant and results from the deflection of the particles in Earth's magnetic field. 
This induces a transverse current in the air shower front, whose strength correlates to the number of particles and their energy, as well as the air density. 
As these two quantities vary as the shower develops, electromagnetic emission in the radio regime is emitted. 
The charge-excess component on the other hand comes from a build-up of negative charge at the shower front. 
Electrons stripped off air molecules travel with the shower front, leaving their positive counterpart behind.
Since this charge imbalance varies over time, it radiates as well.

Most analyses of the radio emission from EAS make extensive use of Monte-Carlo simulations, such as the CORSIKA package \cite{Heck1998} with the CoREAS plugin \cite{Huege2013} for the radio emission.
These microscopic codes are extremely precise but also very compute-intensive, which poses limitations on current analyses. 
This is why we developed template synthesis. 
It is a hybrid simulation approach which makes use of the macroscopic emission components to achieve higher computational efficiency, while still using inputs from microscopic simulations to benefit from their accuracy.

The template synthesis method has been presented in previous work \cite{Desmet2023}, but there it was only applicable to fixed geometries. 
Here we present a generalisation to the framework. 
We have found a set of scalings which allow us to formulate the ``spectral functions'' in a way that does not longer depend on the air shower geometry.
These are extracted from a set of CoREAS simulations with a zenith angle of 50 degrees, but can be applied to showers with lower zenith angles as well. 

\section{The template synthesis method for fixed geometries}
\label{sec:template_synthesis}

\begin{figure}
  \begin{center}
    \includegraphics[width=0.5\textwidth]{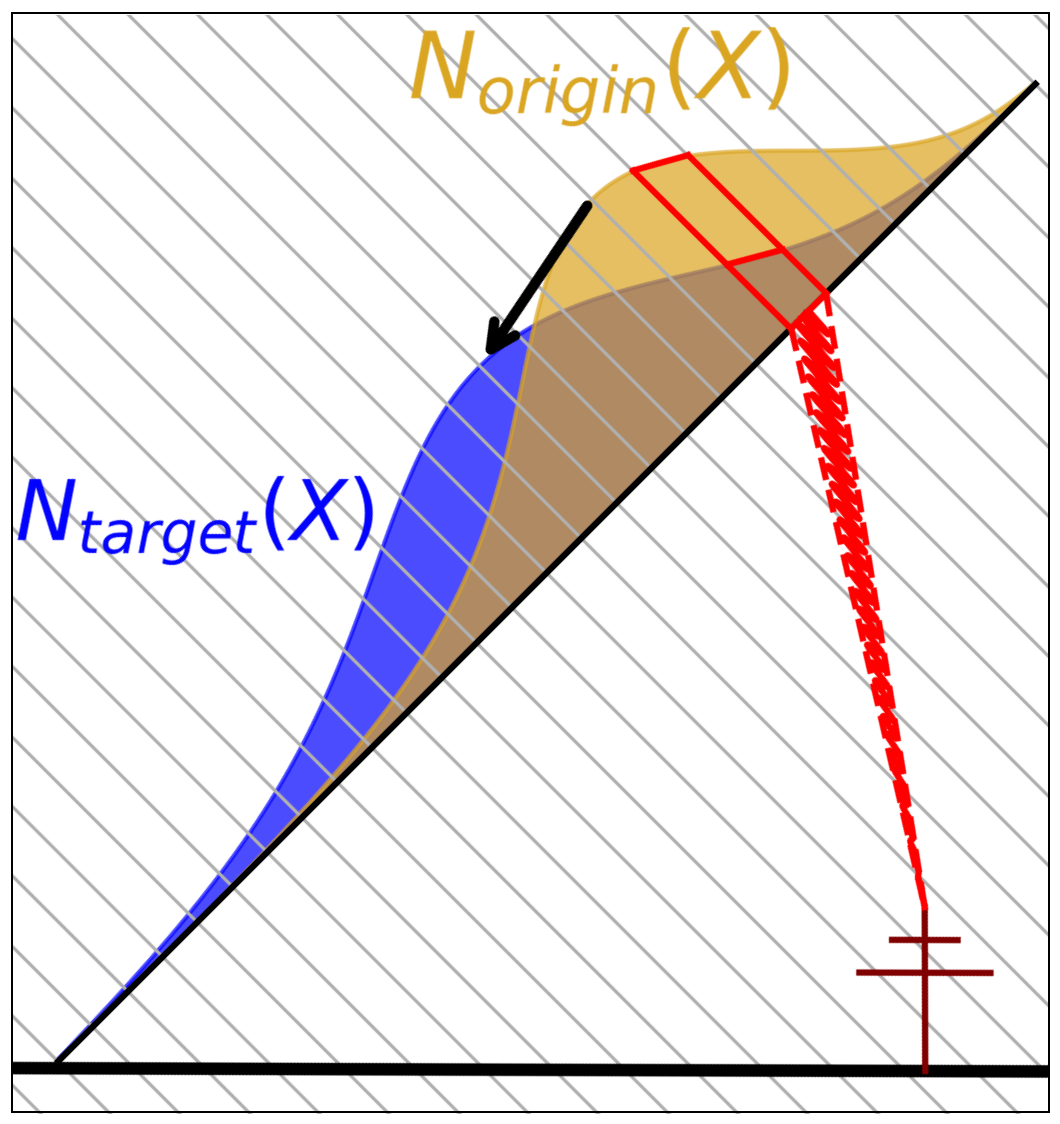}
  \end{center}
  \caption{
    Schematic overview of the slicing in template synthesis. 
    The slanted black line represents the shower axis, with the yellow and blue curves depicting the longitudinal profiles of the origin and target showers respectively. 
    During the synthesis process, we consider the emission coming from each atmospheric slice (one marked as a red box) separately. 
    In order to retrieve the emission from the target shower, we rescale the emission of the origin shower with respect to the target shower properties such as particle number and shower age.
  }\label{fig:slicing}
\end{figure}

The template synthesis method uses a microscopically simulated input shower, called the origin shower, to synthesise the emission from an EAS with any given longitudinal profile \cite{Desmet2023}. 
Core to this approach is the concept of subdividing the atmosphere into slices of constant atmopsheric depth, as illustrated in Figure \ref{fig:slicing}.
In CoREAS, we can configure our simulation to save the radio emission coming from the slices into different files. 
We bin the longitudinal profile in CORSIKA into the same slices, to have the number electrons and positrons per slice.

In order to obtain the parametrisations required for template synthesis, we use a library of CoREAS showers, sliced in atmosphere depth. 
In each antenna, we then process the radio emission coming from every slice separately. 
The first step is to normalise the emission with the number of electrons and positrons in the slice. 
To achieve this, we simply divide the electric field trace by this sum. 
We then split the radio emission into the two macroscopic emission components, the geomagnetic and charge-excess. 
This is possible due to the fact that these have different polarisation patterns. 
By choosing our simulated antenna's on the $\vec{v} \times \vec{v} \times \vec{B}$ axis, we can readily decouple the two components \cite{Glaser2016}.

To both these components we apply a similar procedure. 
We first transform the time traces to the frequency domain, and then fit a function to the amplitude in frequency space. 
The parameters of this function we call the spectral parameters. 
We fit a parabola to each spectral parameter as a function of the \textXmax\ of the shower they were extracted from. 
The resulting function we refer to as the spectral function.
In each antenna, we have one spectral function per slice. 
This function tells us for any given \textXmax\ the expected value of the spectral parameters in that slice, which in turn give us the expected amplitude frequency spectrum.
In other words, they describe how the pulse shape depends on the shower age, when all other parameters (like particle number) are fixed.

To synthesise the emission from any shower, we require the longitudinal profiles of both the origin and target showers, as well as the origin's radio emission per slice. 
In every slice, we correct the amplitude spectrum for the particle number in both profiles. 
We then also correct the spectrum using the ratio of the expected spectra, as given by the spectral functions. 
The phase spectrum is not changed, as the geometrical setup is kept identical (i.e. we synthesise the emission in the same antenna we have microscopically simulated).

We have shown before that using this approach, we can achieve a precision of 6\% in amplitude over a broad frequency range of 20 to 500 MHz \cite{Desmet2023}. 
Still, this is currently limited to the geometries we have the spectral functions for.
To construct these functions we need a large library of microscopically simulated showers, so in order to really benefit from template synthesis we need to find a generalisation of these. 

\section{A geometrical setup based on antenna viewing angle} 
\label{sec:viewing_angle}

What we need in order to synthesise the emission for air showers with any geometry, is a description of the spectral functions which is geometry-independent. 
However, in the current formulation of template synthesis this is hindered due to the fact that we have one spectral function per combination of slice and antenna. 
When trying to compare two geometries, it is not clear which slices are equivalent, as for example the density in a slice at some atmospheric depth changes when changing zenith angle. 
Similarly, we have to understand what the equivalent antenna is when going to a different geometry. 

Therefore we introduce a new variable, which will hold the information of the relation between slice and antenna, namely the \textbf{viewing angle}. 
We express this in units of the local Cherenkov angle of the slice, which is calculated from the refractive index $\slice{n}$ at the vertical height of the slice as
\begin{equation}
  \slice{\theta}^C = \arccos \left( \frac{1}{\slice{n}} \right) \; .  
  \label{eq:cherenkov}
\end{equation}
The viewing angle is defined as the opening angle with respect to the shower axis of the straight line connecting the slice and the antenna's projection along the shower axis into the shower plane. 
The reason for choosing the projection, is that we will later interpolate the signals. 
This interpolation happens in the shower plane, and as such it makes more sense to connect the viewing angle to the position in the shower plane. 
Because projecting an antenna to the shower plane is usually done along the shower axis, the viewing angle of the physical and projected antenna's are generally not equal. 
Though for zenith angles below 50 degrees the difference is smaller than a degree for most slices.

\begin{figure}
  \begin{center}
    \includegraphics[width=0.55\textwidth]{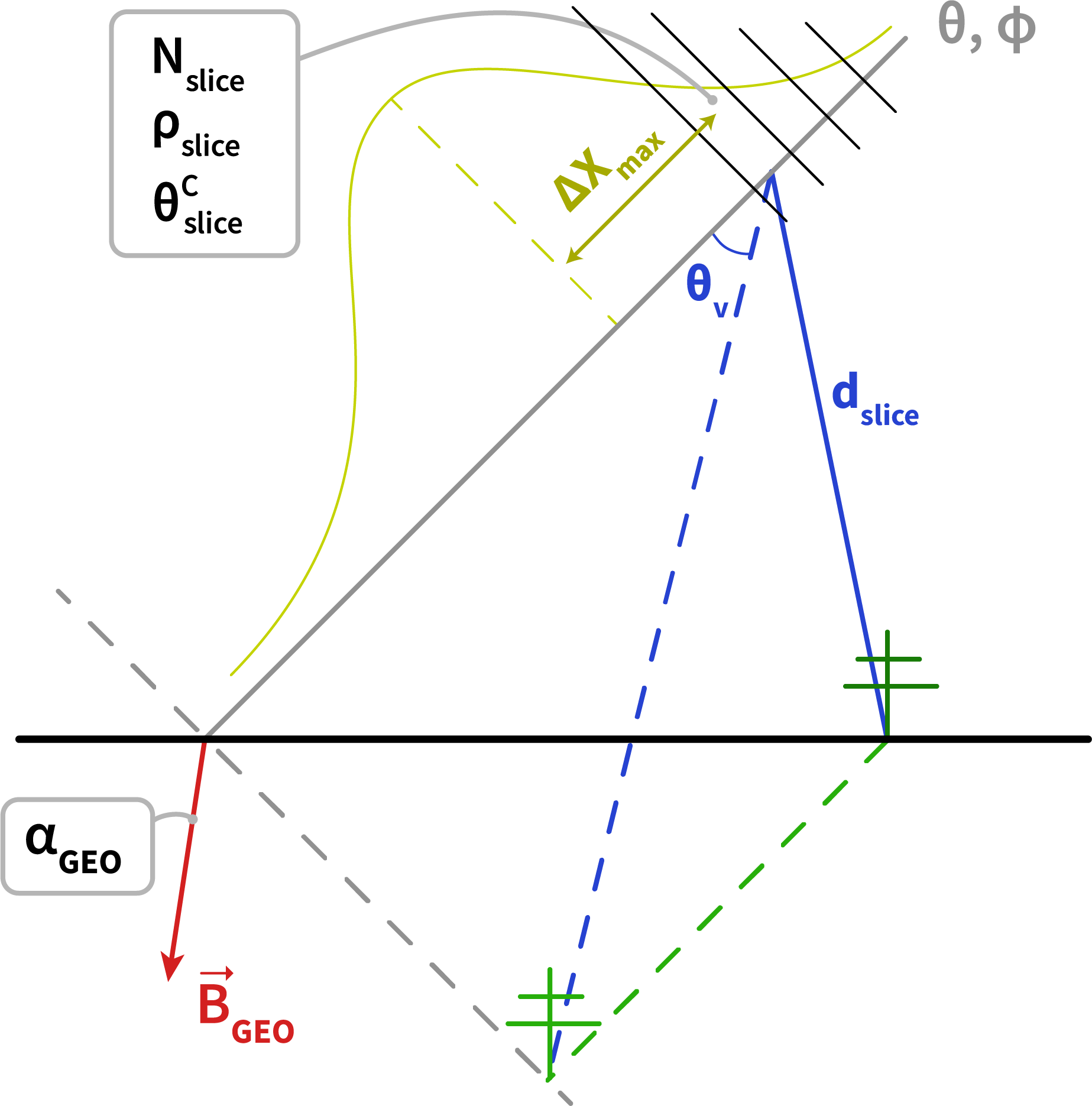}
  \end{center}
  \caption{
    The geometrical setup used for the generalisation of the template synthesis approach.
    We represent the shower axis by the slanted grey line, marked with the shower zenith angle $\theta$ and azimuth angle $\phi$. 
    The geomagnetic angle $\alpha_{\text{GEO}}$ is the angle between this axis and the magnetic field vector $\vec{B}_{\text{GEO}}$.
    For each atmospheric slice we have its properties, the number of particles, the density and the local Cherenkov angle, as well as the distance to \textXmax . 
    An antenna's position with respect to a slice is given by its distance to it, as well as the viewing angle $\theta_V$ under which the antenna projected in the shower plane sees the slice.  
  }\label{fig:viewing_angle_geometry}
\end{figure}

As such we arrive to the setup shown in Figure~\ref{fig:viewing_angle_geometry}. 
The slices are now labelled with their distance to \textXmax\ in \textgcm\ instead of their absolute grammage. 
This comes from our observation that the shower age seems to be the quantity that governs the pulse shape. 
Other factors that will influence the (amplitude) of the emission are the number of emitters in the slice, $\slice{N}$, the density of the slice, $\slice{\rho}$ and the local Cherenkov angle of the slice, $\slice{\theta}^C$.
The distance from the slice to the physical antenna $\slice{d}$ will also play an important role. 
Lastly, for the geomagnetic component we will also have to account for the geomagnetic angle $\alpha_{\text{GEO}}$. 


\section{Template synthesis across geometries} 
\label{sec:Template synthesis across geometries}

\begin{figure}
  \centering
  \begin{subfigure}{0.9\textwidth}
    \includegraphics[width=0.95\textwidth]{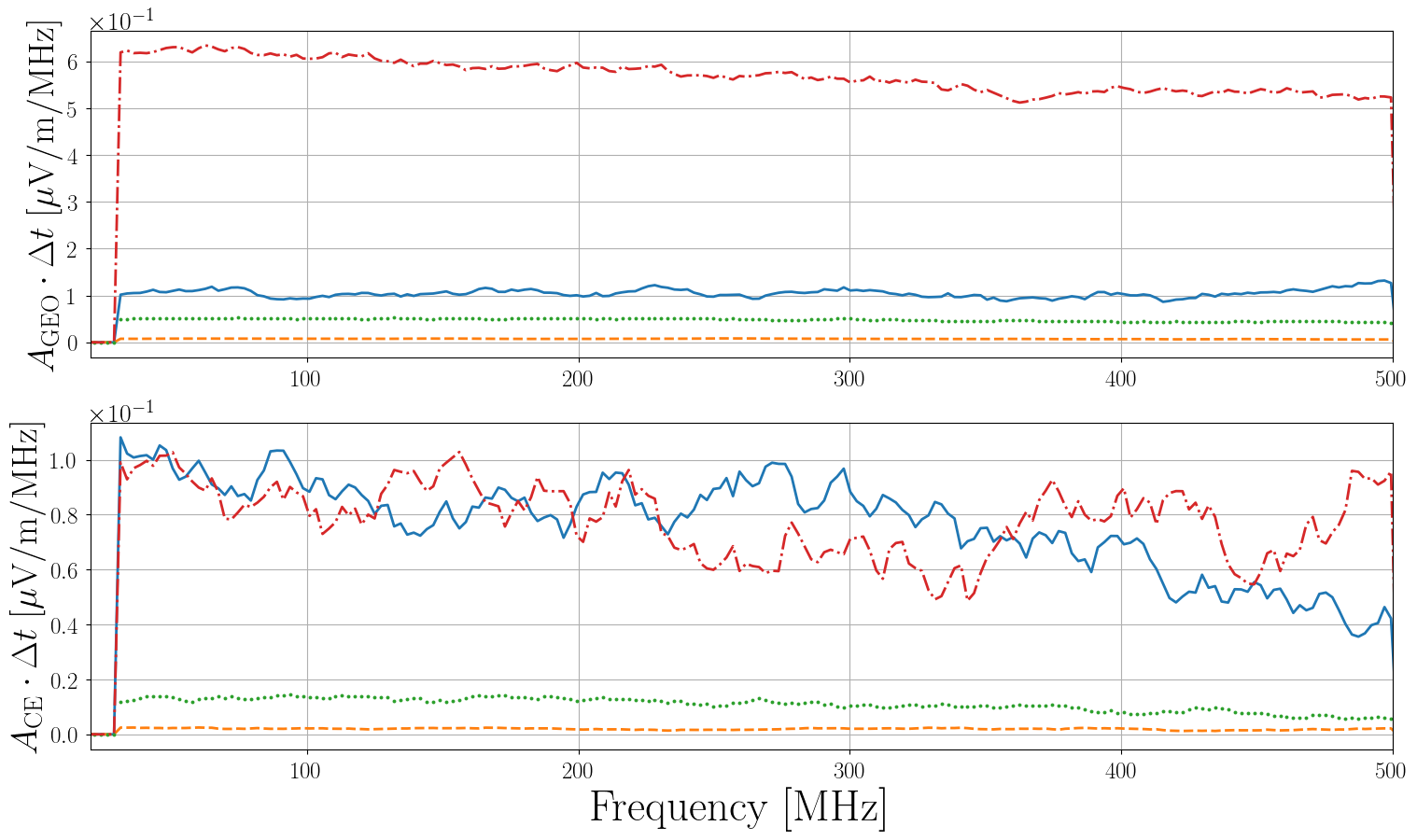}
    \caption{The amplitude frequency spectra from showers with different zenith angles, without any scaling.}\label{fig:spectra_unscaled}
  \end{subfigure}
  \begin{subfigure}{0.9\textwidth}
    \includegraphics[width=0.95\textwidth]{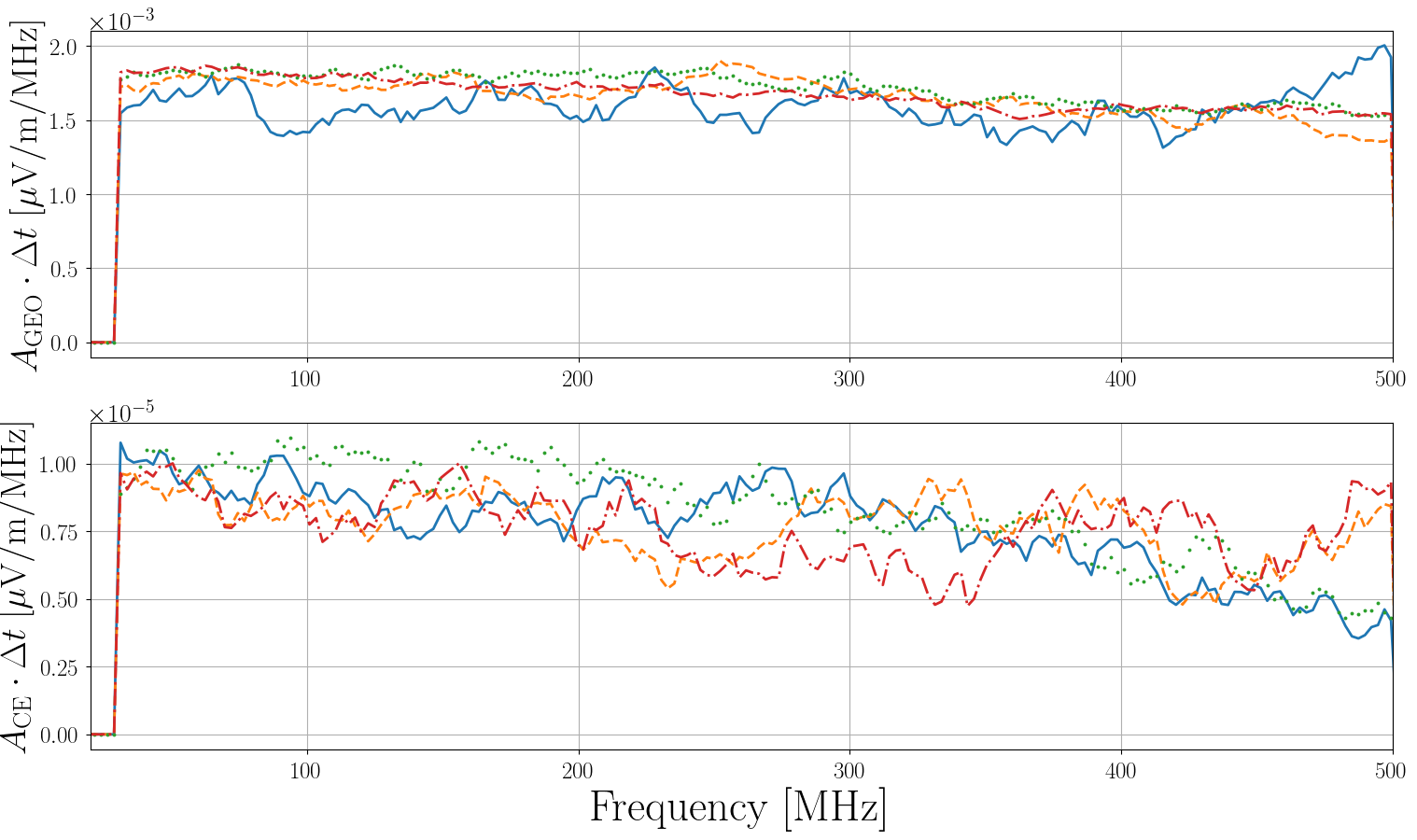}
    \caption{The same spectra as shown in (a), after applying the scaling relations mentioned in the main text.}\label{fig:spectra}
  \end{subfigure}
  \caption{
    We compare the spectra from four different air showers, each with a different zenith angle (while keeping the azimuth angle fixed).
    The solid blue line is for a shower with 30 degree zenith angle, the dashed yellow line corresponds to 40 degree zenith, the dotted green line has a zenith angle of 45 degrees and the dash-dotted red line is from a shower with 50 degree zenith angle.
    From each of them, we look at the geomagnetic (GEO, top) and charge-excess (CE, bottom) amplitude frequency spectra coming from the slice which is 200 \textgcm\ before the \textXmax\ of that respective shower, in an antenna which is looking under a viewing angle of 0.8 times the slice Cherenkov angle. 
    In (a) we see that without correcting for the geometry, the spectra have a seemingly random ordering. 
    Once we apply the scalings mentioned in the main text in (b) however, all spectra overlap and can be described by a single function.
  }
\end{figure}

If we compare the amplitude frequency spectra from slices with the same \textDeltaXmax\ and in antennas looking under the same viewing angle (which, as a reminder, is given in units of the slice Cherenkov angle $\slice{\theta}^C$), we still see significant differences as in Figure \ref{fig:spectra_unscaled}.  
Thus, before we can proceed have to remove this dependency on the geometry. 
The properties mentioned in the previous section will all play an important role. 

In the end we landed on the following set of scaling relations, which are supposed to map the spectra from Figure \ref{fig:spectra_unscaled} onto each other.
\begin{itemize}
  \item The geomagnetic component is scaled with 
  \begin{itemize}
    \item the sine of the geomagnetic angle, $\sin\left( \alpha_{\text{GEO}} \right)$ and 
    \item the inverse of the air density, $\rho^{-1}_{\text{slice}}$ .
  \end{itemize}
  \item The charge-excess emission on the other hand scales with 
  \begin{itemize}
    \item the sine of the local Cherenkov angle, $\sin\left( \theta^C_{\text{slice}} \right)$ .  
  \end{itemize}
  \item Both components also scale with 
  \begin{itemize}
    \item the inverse of the distance from slice to antenna, $d^{-1}_{\text{slice}}$ and 
    \item the number of emitting particles in the slice, $\slice{N}$.
  \end{itemize}
\end{itemize}
The scalings with air density and with the Cherenkov angle were found in reference \cite{Ammerman_Yebra_2023}.
Applying these scalings to the spectra we had before, we indeed see in Figure \ref{fig:spectra} that all spectra overlap.

From here we can apply the same principles as outlined in Section \ref{sec:template_synthesis}. 
We now describe our spectral functions as a function of \textDeltaXmax , one for every viewing angle that we consider. 
With these in hand, we can synthesise the emission per slice for every viewing angle. 

As the viewing angle is expressed as a fraction of the slice Cherenkov angle, any viewing angle will point to a different distance in the shower plane for each slice. 
In order to retrieve the emission for a given antenna on the ground, we apply the Fourier-based interpolation from \cite{Corstanje_2023} in every slice.

\begin{figure}
  \begin{center}
    \includegraphics[trim={0 0 0 2cm},clip,width=0.94\textwidth]{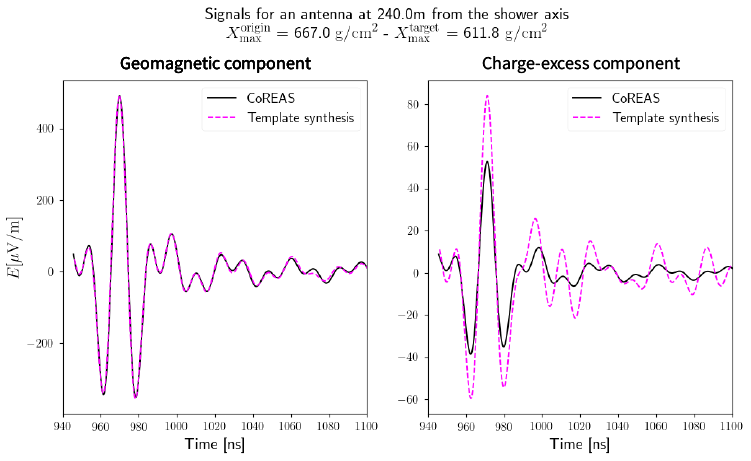}
  \end{center}
  \caption{
    Here, we show the radio emission in an antenna at 240\,m from the shower axis, as simulated by CoREAS (black) and synthesised by the generalised template synthesis method (magenta).
    The origin shower has an \textXmax\ of 667 \textgcm , while the target shower has an \textXmax\ of 611.8 \textgcm , which is a difference of about 55 \textgcm . 
    Both showers have a zenith angle of 50\textdegree , which is also the zenith angle of the showers used to construct the spectral functions.
    The signals in this plot are filtered in the [30, 80] MHz frequency band. 
    We can see that for the geomagnetic component the synthesised signal matches very well with the CoREAS simulated trace, but for the charge-excess component template synthesis overestimates the signal. 
  }\label{fig:signal}
\end{figure}

Once the emission from each slice is synthesised, we can sum them all up to obtain the complete signal in the antenna. 
This can then be compared to a CoREAS simulation, as shown in Figure \ref{fig:signal}. 
Here we show the synthesised signal filtered in the 30 to 80 MHz band in an antenna at 240\,m away from the shower axis.
We see that we already synthesise the geomagnetic component very well, but the charge-excess emission is showing a stronger signal than expected. 
We suspect it might have to do with the interpolation.
It is known that the interpolation method can have some issues if the signals are too weak, and the charge-excess is indeed the smallest of the two.
Therefore, we will move to a different interpolation scheme in the near future.
Instead of interpolating the synthesised electric fields, we will interpolate the value of the amplitude frequency spectra coming from the spectral functions.
Our first results using this approach indicate a much improved performance for the charge-excess component.

\section{Conclusion} 
\label{sec:Conclusion}

In this contribution we presented a generalisation of the template synthesis framework. 
Our goal was to find a description of the spectral functions which does not longer depend on the air shower geometry. 
Once we have these, we can apply the synthesis procedure to any microscopic, sliced air shower simulation. 
To achieve this, we reformulated our geometrical setup. 
The atmospheric slices are labelled by their distance the shower maximum, in \textgcm , allowing us to drop the dependency on the slice in the spectral functions. 
Using the slice \textDeltaXmax\ in this way makes the dependency of the pulse shape on the shower age explicit.

Instead of having a spectral function per pair of slice and antenna, we now have one function per viewing angle which depends on the slice \textDeltaXmax . 
The viewing angle is defined as the opening angle between the shower axis and line-of-sight from the slice to antenna projected into the shower plane. 
Crucially we express this angle as a fraction of the local slice Cherenkov angle, which depends on the local refractive index. 
When comparing amplitude spectra from showers with different zenith angles, we noticed a dependency of the spectral parameters on the air shower geometry. 
However, we found a set of scaling relations which correct for this. 

Once we extracted the general spectral functions, we synthesised the signal in an antenna and compared the trace to the one coming from a CoREAS simulation. 
While the geomagnetic component was synthesised with excellent accuracy, the charge-excess component was strongly overestimated. 
We suspect this might be related to the interpolation procedure which is performed on a per-slice basis. 
As the signals can become really weak for some slices, the interpolation might struggle in those. 

In order to mitigate this, we plan to integrate the interpolation approach more directly into template synthesis in a forthcoming publication. 
By interpolating the frequency spectra calculated using the spectral functions, we avoid instabilities that can occur when applying the interpolation to the small electric field amplitudes coming from a single slice.
We will also change our approach to projecting antenna's to the shower plane, by using the line-of-sight vector instead of the shower axis.
While we do not expect this to have a big impact, it does remove the ambiguity mentioned before, where the physical and projected antenna do not have the same viewing angle.


{\bf Acknowledgment}
This research is supported by the Flemish Foundation for Scientific Research (FWO-AL991).

\bibliographystyle{JHEP}
\bibliography{sources}

\end{document}